\documentclass{article}
\usepackage{amsmath, amssymb, amscd}
\usepackage{setspace}
\date{\empty}
\author{Fouad B. Chedid\footnote{A'Sharqiyah University, Ibra, Oman, f.chedid@asu.edu.om}}

\title{Kolmogorov Complexity and Information Content}
\begin{document}
\maketitle
\begin{abstract}
In this paper, we revisit a central concept in Kolmogorov complexity in which one would equate program-size complexity with information content. Despite the fact that Kolmogorov complexity has been widely accepted as an objective measure of the information content of a string, it has been the subject of many criticisms including the fundamental one directed by logicians and philosophers towards the statistical and semantical theories of information, which is about confusing an object and its name. In this paper, we clarify a number of subtle issues that are at the center of this debate. 
\end{abstract} 
\section{Introduction}
In his seminal paper (Kolmogorov 1965), Kolmogorov introduced an algorithmic approach to quantitatively define information using the theory of algorithms. In particular, the Kolmogorov complexity of a finite binary string $x$ is defined as the length of a shortest binary program $p$, which when interpreted by some Universal Turing Machine $M$, causes that machine to output $x$ and then halt.   Formally put,
\begin{align*}
K_M(x) &= \min._{p}\{|p|: M(p)=x\}\\
&= \infty, \text{ if there is no }p\text{ such that }M(p)=x. 
\end{align*}
So, one can view $K_M(x)$ as the size of an ultimate, most compressed, code $p$ for $x$ using a universal Turing machine as a decompressor. Unfortunately, the compressor that generates $p$ is not known. This is true because of the Halting problem (Turing 1936).  That is, there is no effective procedure for determining the size of a smallest program for $x$. Of course, we can always estimate $K_M(x)$ from above by discovering shorter and shorter programs for $x$, but we can never be certain that our best discovered short program for $x$ is a shortest one.  

In that same paper (Kolmogorov 1965), Kolmogorov used the expression "the information content of a message", and then he went on to define $I(x:y)$ as "the quantity of information conveyed by an individual object $x$ about an individual object $y$", and concluded that the information conveyed by $x$ about itself, $I(x:x)$, is equal to $K_M(x)$. 

We mention that prior to this work of Kolmogorov, the notion of the information content of a string wasn't there. For example, Shannon's work (Shannon 1948) was about the minimum number of bits needed on the average to transmit a value taken by a random variable, as a syntactic unit independent of any semantic (Shannon wrote `` ... semantic aspects of communication are irrelevant to the engineering problem ..''). Similarly, Chaitin was interested in studying the size of a shortest binary program capable of generating a given sequence of bits on a universal Turing machine (Chaitin 1966), and Solomonoff was interested in predicting the next value taken by a random variable following an unknown probability distribution (Solomonoff 1964). 

The following definition of a possible information measure was first suggested by Wiener: ``The amount of information provided by a single message $m_i$, $I(m_i) = -\log_2 p_i$'' (Wiener 1948), which is related to the number of bits needed to identify any of the messages which happen to occur with probability $p_i$. Still, we should be clear that neither Wiener nor Kolmogorov has made the shift from the sense of treating information as a sequence of bits to the sense where information is about what is expressed by that sequence of bits, which was the main drive behind what has become known as the semantic theory of information (Popper 1934, Bar-Hillel 1952, Carnap 1964). The comforting thing is that both statistical (= Shannonian) and semantical theories of information agree that {\em information is about removal of uncertainty}. This agrees with the point of view suggested by Kolmogorov, though seen from an opposite end, that information is about uncompressed regularities. That is, the more compressible a string is, the less information it contains, and vice versa. For a wealth of information on Kolmogorov complexity and its applications, the reader is referred to Cover and Thomas 2006 and Li and Vit\'anyi 2008. 

Despite the fact that Kolmogorov complexity has been widely accepted as an objective measure of the information content of a string, the concept of equating program-size complexity with information content continues to be challenged using several arguments (See for e.g., Grassberger 1990,  Li 1991, Gell-Mann and Lloyd 2003) including the fundamental criticism directed by logicians and philosophers towards the statistical and semantical theories of information (Raatikainen 1998), which is about confusing an object and its name, or equivalently confusing the object language and the metalanguage.  We show that understanding Kolmogorov's notions of randomness in the positive versus negative sense is at the center of this debate. We also explain the works of Shannon, Solomonoff, and Kolmogorov using the language of analytic philosophy. We clarify that both Shannon's entropy and Solomonoff's universal probability {\em use} their arguments and that Kolmogorov complexity {\em mentions} its argument, which explains the suitability of Kolmogorov complexity as a measure of the information content of an individual string.

The rest of the paper is organized as follows. Section 2 gives the necessary background about related theories. We discuss the notions of Shannon's entropy, Solomonoff's universal probability, and Kolmogorov's algorithmic description.  Section 3 contains our contribution. 
\section{Background}
Shannon's information theory is about the most economical way of transmitting a message between a sender and a receiver over a binary communciation channel. For simplicity, we assume that messages are the outputs of a discrete random variable $X$ that can take its values from a finite alphabet $\mathbb X=\{x_1, \ldots, x_n\}$ using  a probability distribution $P=\{p_1, \ldots, p_n\}$. That is, Prob.$\{X=x_i\}=p_i, \forall 1\le i\le n$. We also say that $X\text { is distributed according to } P(x)$.

We mention that the Shannon-Fano code (Shannon 1948, Fano 1949) is the most fundamental result in determining the descriptive complexity of an object in both probabilistic (Shannon's) and algorithmic (Kolmogorov's) settings. It is a simple observation which states that if an event has a probability of occurrence $p$, then the number of possible outcomes of that event is $\frac 1 p$, and the smallest number of bits needed to represent any of those outcomes is $\lceil \log \frac 1 p\rceil$. Then, it is a straighforward result that the expected value of the smallest number of bits needed to represent any of the outcomes of the random variable $X$ is $$\sum_{x_i\in \mathbb X}p_i\lceil \log \frac 1 p_i\rceil.$$
Shannon had wanted to call this quantity the ``uncertainty'' or ``unexpectedness'' of the random variable, but then he went along with a suggestion from John von Neumann and calledl it ``entropy''\footnote{In thermodynamics, if we let $n$ be the number of ways in which a physical sysem $S$ can be configured, then the quantity defined by $\log n$ is a measure of uncertainty or disorganization in $S$, also called entropy of $S$.} (Wicken 1987). The entropy of a random variable $X$ drawn according to a mass probability function $p$ is often denoted by $H(X)$ or $H(p)$. The idea is that when a receiver learns about some output value $x_i$, then by doing so, she eliminates $2^{H(X)}$ possible outcomes on the average, which is a measure of the average unexpectedness or uncertainty stored in $X$. Furthermore, if we accept the notion that removal of uncertainty constitutes information gain, then the entropy of the random variable becomes a measure of the average amount of information stored in $X$. 

In the Shannon's theory of information, one of the main issues was to find the length of a shortest description, on the average, of a string of symbols given those symbols' probabilities of occurrence. We now know that the solution to this specific issue is the entropy of the probabitliy distribution $p$ of the symbols given by $H(p) = -\sum_x p_x\log p_x$. Thus, the entire theory of Shannon rests on the assumption that the probability distribution of the information source is made available to us in advance.  

We next ask what happens if we don't know the probability distribution of the information source? This is the same question that Solomonoff asked himself in the early 1960s, although his intention was to develop a formal theory of inductive inference. Investing in the main idea of Wiener and Shannon on the relationship between the length $l_x$ of an optimal code for a symbol $x$ and the probability of occurence $p_x$ of $x$; that is, $l_x=-\log p_x$, or equivalently, $p_x=2^{-l_x}$, Solomonoff combined ($i$) the simple idea that whatever binary encoding the sender develops for $x$, it is something that the receiver must be able to decode in order to recover $x$ and ($ii$) the widely accepted Church's thesis about the Turing Machine as being the most general model capturing the notion of effective computation to propose that whatever encoding $e_x$ of length $l_x$ the sender decides to use to encode $x$, the receiver (Turing machine) should be able to recover $x$ from $e_x$. In particular, if we let $M$ be a fixed Universal Turing Machine, then we want $M(e_x) = x$. In this sense, $e_x$ acts as a program that when interpreted by $M$ causes $M$ to generate $x$ and then halt. Moreover, we already know that there is a relationship between $l_x$ and the probability distribution generating $x$ as assumed by the sender, namely $p_x=2^{-l_x}$. Putting all these ideas together, Solomonoff proposed to model the true probability distribution of $x$, whatever that may be, by his notion of Universal Probability\footnote{Taking all programs for $x$ into account follows Epicurus' approach, while limiting our attention to the shortest program for $x$ follows Occam's razor.} defined as
$$ P(x) = \sum_{p: M(p)=x}2^{-|p|}$$
which is basically the probability that a random string will print out $x$. An additional requirement is needed here for $P(x)$ to be a true probability. $P(x)\le 1$ if and only if all programs $p$ such that $M(p)=x$ form a prefix-free set\footnote{Prefix-freeness means that if $M(p)=x$, then there is no other prefix $q$ of $p$ $(q\not= p)$ that can cause $M$ to print $x$ and then halts.}. This follows from Kraft Inequality (Shannon 1948, Kraft 1949).
\vskip.1in
Around the same time in the early 1960s, Kolmogorov developed similar ideas to Solomonoff's, although for the purpose of studying the algorithmic complexity of randomness. To motivate this discussion, let us first consider an experiment in which we flip a fair coin 30 times and write down $1$ if the coin shows Head and $0$ if it shows Tail. Let $x_1$ and $x_2$ be the outputs of two runs of this experiment. 
$$ x_1: 101010101010101010101010101010$$
$$ x_2: 110001010110100010111011101010$$
We next ask which of these two strings is the more random? Although $x_1$ and $x_2$ have the same probabilities of occurrence $(= 2^{-30})$, it is obvious that these two strings have different levels of randomness with $x_2$ being the more random. This is true because while $x_1$ consists of 15 repetitions of the pattern``10'', there doesn't seem to be an obvious way to generate $x_2$ than by rewriting $x_2$ in its entirety, and so, we would like to conclude that $x_2$ is more random than $x_1$. As a matter of fact, Laplace (1749-1827) explained this a long time ago by considering the set $S$ of all binary strings of length $30$, and by showing that among all elements of $S$, the size of the subset $S_1\subset S$ of which $x_1$ is a typical element is much smaller than the size of the subset $S_2\subset S$ of which $x_2$ is a typical element, and hence, the probability of the event $S_1$ is smaller than that of $S_2$, which explains why $x_1$ is less random than $x_2$.

Kolmogorov, similar to Solomonoff, used the widely accepted Church's thesis about Turing machines to propose an algorithmic description of a binary string $x$ as the length of a shortest program $x^*$ for $x$ on a universal Turing machine. In particular, if we let $M$ be a fixed Universal Turing Machine, then we want $M(x^*) = x$. In this sense, $M$ acts as a decompressor of an ultimately compressed program (encoding) for $x$. Kolmogorov called the length of $x^*$ the descriptive complexity of $x$ (also known as Kolmogorov complexity) to be denoted by $$K_M(x) = |x^*|$$ 
or equivalently, $$K_M(x) = min_p\{|p|: M(p)=x\}.$$
We note that the Kolmogorov Complexity of a string is defined up to an addtive constant. This is true because on one hand, $K_M(x)$ is defined relative to a fixed Universal Turing Machine and on another hand we have the Invariance Theorem, which states that given two fixed universal Turing machines $M_1$ and $M_2$, $|K_{M_1}(x) - K_{M_2}(x)|\le c$, for some positive constant $c$ which depends on $M_1$ and $M_2$, but not on $x$. This is why it is fair to say that (up to an additive constant) the Kolmogorov Complexity of a string is an intrinsic property of $x$ itself and is independent of the fixed universal Turing machine used to generate it. Hereafter, we drop the reference to any fixed universal Turing machine and write $K(x)$ for the Kolmogorov Complexity of $x$.

Knowing that there is a relationship between the length of an encoding for $x$ and the probability of occurence of $x$, we have
$$Prob.(x)=\sum_{x^*}2^{|K(x)|}.$$
 An additional technical requirement is needed here for $Prob.(x)$ to be a true probability. $Prob.(x)\le 1$ if and only if all shortest programs $x^*$ for $x$ form a prefix-free set\footnote{This is obviously true because otherwise there would be some shortest program that is shorter than another shortest program, which is a contradiction.}.

Comparing the expression of $Prob.(x)$  to Solomonoff's universal probability shows that Kolmogorov's work supports the principle of Occam's razor\footnote{William of Ockham (1287-1347):  If a phenomenon can be explained using several theories, then it is best to adopt the simplest (shortest?) one.} while Solomonoff's work supports the principle of Epicurus\footnote{Epicurus (341-270 BC): If a phenomenon can be explained using several theories, then it is important to keep all of them.}. Solomonoff believed it is important to consider other programs (other than a shortest one) for $x$ because by not doing so, we would be missing on some information that could be helpful for our understanding of the true distribution of $x$ (Personal communication 2008). 

Kolmogorov writes ``In practice, we are most interested in the quantity of information conveyed by an individual object $x$ about an individual object $y$'' (Kolmogorov 1965). The conditional (relative) Kolmogorov complexity of a string $x$ when another string $y$ is given is 
$$ K(x/y) = \text{min.}_p\{|p|: M(y,p) = x\}$$  
Here, we follow Kolmogorov's notation and place the auxilliary information $y$ before $p$. Kolmogorov argued that since $K(x/y)\le K(x)$, it is fair to call the differene $K(x)-K(x/y)$ the amount of information in $y$ about $x$, and so, he defined
 $$I(y:x) = K(x) - K(x/y)$$
as the amount of information in $y$ about $x$. If we next compute $I(x:x)$, we have $K(x/x) = \text{min.}_p\{|p|: M(x,p) = x\} = O(1)$, where the $O(1)$ term is the length of the copy program $p$ that copies its input to its output. Thus, up to an additive constant, $K(x/x) = 0$ and $I(x:x)=K(x)$. Thus, Kolmogorov suggested to call $K(x)$ the amount of information in $x$ about itself\footnote{We could as well call it self-information, similar to the notion of self-entropy in Shannon's information theory.}. 

In the next section, we show that understanding Kolmogorov's notions of randomness in the positive versus negative sense is at the center of this debate. We also explain the works of Shannon, Solomonoff, and Kolmogorov using the language of analytic philosophy. We clarify that both Shannon's entropy and Solomonoff's universal probability {\em use} their arguments and that Kolmogorov complexity {\em mentions} its argument, which explains the suitability of Kolmogorov complexity as a measure of the amount of information in an individual string.
\section{Criticisms and Clarifications}
\subsection{Equating Program-Size Complexity With Information Content}
According to Kolmogorov complexity, the information content of a string is solely determined by the size of a shortest program capable of generating that string on a Universal Turing Machine. Equating program-size complexity with  information content implies that strings that admit short descriptions (or equivalently, that can be significantly compressed) have low complexity (little information) and strings that resist short descriptions have high complexity (lots of information). This equality is an ingenious idea that, while supported by many, has been the subject of many criticisms. To motivate this discussion, consider, for example, the case of two strings $x$ and $y$ of 1000 characters each, where $x$ is a passage from Tolstoy's masterpiece War and Peace and $y$ is a string of 1000 characters chosen at random. Using Kolmogorov complexity as a measure of information content, we conclude that $x$ has less information than $y$. This is true because it is always possible to encode the string $x$ by abbreviating some sentences in it to be recovered by the reader (say we use ``btw'' for ``by the way'', etc.). However, the only way to generate $y$ is to copy it in its entirety. This absurdity is a main argument used against using program-size complexity as a measure of information content. For example, the German physicist Grassberger writes ``Kolmogorov complexity seems to be the quantity most closely related to the intutive notion of randomness, not of complexity'' (Grassberger 1986). Grassberger suggested using the word ``randomness'' for the property measured by Kolmogorov complexity and reserving the word ``complexity'' for something that is between total regularity and total randomness (Grassberger 1990). Thus, according to Grassberger, strings of total regularity (those that can be significanlty compressed) or total randomness (those that can't be compressed at all) have small complexity. In either case, the string can be generated by a simple procedure. Here, simplicity is not related to the size of the procedure, but rather to the process involved in the generation of the string. Similarly, the American physicist Gell-Mann supported the unsuitability of the quantity measured by Kolmogorov compelxity as a measure of complexity (Gell-Mann 2003). Furthermore, Gell-Mann defined the Effective Complexity of an object as the length of a highly compressed description of that objects' {\it regularities} (Gell-Mann 1996). Another interesting view was proposed in (Li 1991), from which we quote the following text:
\begin{quote}
It has been pointed out repeatedly that algorithmic complexity does not fit our intuitive notion of complexity. One vague argument is that algorithmic complexity measures the difficulty of the task -- to reconstruct a sequence exactly -- that is not very ``meaningful.'' To say that this task is not very meaningful is partly based on our daily experience that we can pick up the main feature of a picture or the basic meaning of a text by quickly scanning it without paying attention to every detail. A similar argument for algorithmic complexity not being a ``good'' measure of compelxity is that the effort put in to generate a sequence to be exactly correct does not contribute to the ``important features'' of that sequence. Rather, they contribute too much to the ``trivial'' details.
\end{quote}
We now return to Kolmogorov and recall some of his work from 1974. Kolmogorov (though he never published anything on this himself) proposed a structure function $h_x()$ (Cover 1985) as a way to divide a shortest program $x^*$ for a finite binary string $x$ into two parts responsible for encoding the regular and nonregular (random) aspects of $x$ separately. Of course, both parts are maximally random, and hence it is not clear how such a division can be made. Still, the clever idea is that while we agree that the length of a shortest program $x^*$ for $x$ is a measure of the amount of information (let's call it total information) in $x$, we would like to know more about the constituents of $x^*$. In particular, we would like to know how much of $x^*$ is responsible for coding the regular aspects of $x$ (to be associated with meaningful information in $x$) and how much is responsible for coding the random aspects of $x$ (to be associated with accidental information in $x$). Kolmogorov suggested to model the regular aspects of $x$ by a finite set $S$ of which $x$ is a typical element. The notion of Kolmogorov's minimal sufficent statistic as the length of an optimally compressed description of the regular aspects of a string in the model class of finite sets was then coined. A string $x$ for which its minimal sufficient statistic is about the length of $x$ was called absolutely nonrandom (nonstochastic) by Kolmogorov. Thus, an absolutely nonrandom string contains almost no random information. In other words, almost all regularities in a nonrandom string are incompressible. Kolmogorov called such objects {\em random in the negative sense}. Interestingly (and this can be a source of confusion), the incompressibility property of absolutely nonrandom objects is shared with Kolmogorov random objects, which are incompressible because they contain few regularities. Kolmogorov called these objects {\em random in the positive sense}.  We now know that absolutely nonrandom objects exist (See lemma by G\'acs in (G\'acs {\em et al.} 2001), which was first proved by Levin in 1973 (unpublished)). In 2006, Vit\'anyi developed the theory of recursive function statistic and explained the importance of using a total recursive function to model the regular aspects of an object  if we are to be able to identify absolutely nonrandom objects (Vit\'anyi 2006). This is true because in the model class of partial recursive functions, all objects have low minimal sufficient statistic, the main reason being that the complexity of the model part can always be shifted to the data-to-model part of a universal Turing machine of constant complexity. 

We conclude this section by making the following three observations, which should clarify a lot of confusion as to how the literature on Kolmogorov complexity ought to be read and understood.  
\begin{itemize}
\item It has been common in the Kolmogorov complexity literature to talk about the Kolmogorov complexity of an object on the assumption that there is a one-to-one mapping between objects $x$ and binary specifications $spec(x)$. Unfortunately, the literature doesn't distinguish explicitly between a binary specification and the object it specifies, and this has been a source of confusion for many. If we are to be more explicit, we shouldn't be talking about the Kolmogorov complexity of an object $x$, but about the complexity of the binary encoding, $spec(x)$, of a specification of $x$, and write $ K(spec(x)) = \text{min.}_p\{|p|: U(p)=spec(x)\}$, where $U$ is a fixed reference universal Turing machine. By the same token, when the literature talks about the information content of an object, it is the information content of the binary encoding of a specification of that object that should be inferred. 
\item The Kolmogorov complexity of a finite string $x$ is an objective measure of the amount of information in $x$, which includes information about both regular and random parts in $x$. What one would usually mean by the word ``information'' in ordinary language is really about the amount of information in the regular parts of $x$, which is captured by Kolmogorov algorithmic sufficent statistic, Gell-Mann's effective complexity, or Vit\'anyi's meaningful information. Thus, if we consider a 2-part description $(p,d)$ of a shortest program $x^*$ for $x$, where $p$ and $d$ describe the regular and random parts in $x^*$, respectively, then the amount of meaningful information in $x$ is about the size of $p$, and not the size of $x^*$. Returning to our example at the beginning of this section about the two 1000-character strings $x$ and $y$, where $x$ is a passage from the book War and Piece and $y$ is a sequence of randomly generated characters, let the 2-part descriptions of the shortest progams for $x$ and $y$ be $(p_{x^*},d_{x^*})$ and $(p_{y^*},d_{y^*})$, respectively. It is true that $K(y)>K(x)$, but it is also true that $K(x) \approx |p_{x^*}|$ and $K(y) \approx |d_{y^*}|$. In other words, most of the information content of $x$ is about the regular parts in $x$, while most of the information content of $y$ is about the random parts in $y$. This explains why $x$ contains more meaningful information than $y$. 
\item Kolmogorov distinguished between two types of random strings. Those that are random in the positive sense and those random in the negative sense. While both types of random strings have maximal Kolmogorov complexity, only random strings in the negative sense are complex and have maximal (meaningful) information content. On the other hand, random strings in the positive sense are not complex at all and have little or no information content. An example of a random object in the negative sense would be the United States tax code since it contains many rules and each rule is a regularity. An example of a random object in the positive sense would be a string of 1000 characters selected at random. 
\end{itemize}
\subsection{Confusing an Object and Its Name}
The {\em use-mention} distinction in analytic philosophy is about confusing an object and its name. We shall make this distinction clear using the following, rather peculiar, example:
\vskip.1in\noindent
\indent\indent{\bf {\em Steve has five letters}}.
\vskip.1in\noindent
If this sentence is telling us that the word Steve consists of 5 letters, then we say that this sentence {\em mentions} the word Steve. In here, the word Steve refers to its own text. However, if this sentence is telling us that a man named Steve has received 5 letters in the mail, then we say that this sentence {\em uses} the word Steve. In here, the word Steve refers to a specific person. Thus, {\it mentioning} an object refers to the name of the object, while {\it using} an object  refers to the object itself.
 
We note that confusing an object and its name is a classical philosophical fallacy (Raatikainen 1998). This is because determining the truth value of a sentence depends very much on whether a sentence {\em uses} or {\em mentions} something. For example, in the above example, the {\em mention} of Steve is a true statement because there are indeed 5 letters in the word Steve. However, the truth value of the {\em use} of Steve in the above sentence is not clear at all. It depends on whether Steve has actually received 5 letters in the mail or not. Things can get even more complicated when a sentence is self-referential, which could lead to what is known as the semantic paradox. For example, consider the following statement:
\vskip.1in\noindent
\indent\indent {\bf {\em This statement is false}}.
\vskip.1in\noindent
 The {\em use} of the word This in the above statement could be either true or false, depending on the truth value of the thing This refers to. However, the {\em mention} of the word This gives a contradiction. This is true because if the statement ``This statement is false'' is true, then it cannot be true because it says about itself that it is false. Similarily, if the statement ``This statement is false'' is false, then it must be true. In either case, we get a contradiction. 

Using the language of analytic philosophy, we can say that both Shannon's entropy $H(X)$ and Solomonoff's universal probability $P(x)$ {\em use} their arguments. This is true because both $H(X)$ and $P(x)$ report information about the source probability distributions of their arguments. Thus, the values of $H(X)$ and $P(x)$ are useful only in so far as they describe the true probability distributions of the information sources of their arguments.  

Returning to Kolmogorov complexity, we say that $K(x)$ {\em mentions} $x$. This is because $K(x)$ reports the length of a shortest program capable of generating $x$ on a Universal Turing machine, which is an intrinsic property of $x$. Thus, while both $H(X)$ and $P(x)$ refer to information sources, $K(x)$ refers to the name $x$. This explains why Kolmogorov complexity is a suitable objective measure for the amount of information stored in the individual string $x$. 

Furthermore, it is known that Shannon's entropy $H(X)$ and Solomonoff's universal probability $P(x)$ are closely related to Kolmogorov complexity $K(x)$ (Cover 2006). In particular, we know that 
$$ K(x) \approx  -\log P(x)$$ and $$ H(X) \approx \text{ The expected value of }K(x) $$ 
With this in mind, it is only fair that all three measures be accepted as objective measures for the amount of information in an individual string.

\end{document}